\pdfoutput=1
\documentclass[aps,showpacs,prb,floatfix,twocolumn,amsmath,superscriptaddress]{revtex4-1}
\usepackage{graphicx}
\usepackage{amsmath}
\usepackage{amssymb}
\usepackage{bm}
\usepackage{dcolumn}
\usepackage{subfigure}
\usepackage{units}
\usepackage{hyperref}
\usepackage[usenames]{color}
\hypersetup{pdfborder=0 0 0,colorlinks=true,citecolor=blue,linkcolor=blue}
\input epsf

\begin{document}
\author{Deniz \c{C}ak{\i}r}
\email{deniz.cakir@uantwerpen.be}
\affiliation{Department of Physics, University of Antwerp, Groenenborgerlaan 171, B-2020 Antwerpen, Belgium.}
\author{Diana M. Ot\'{a}lvaro}
\email{d.otalvaro@utwente.nl}
\affiliation{Computational Materials Science, Faculty of Science and Technology and MESA+ Institute for Nanotechnology, University of Twente,
P.O. Box 217, 7500 AE Enschede, The Netherlands}
\author{Geert Brocks}
\email{g.h.l.a.brocks@utwente.nl}
\affiliation{Computational Materials Science, Faculty of Science and Technology and MESA+ Institute for Nanotechnology, University of Twente,
P.O. Box 217, 7500 AE Enschede, The Netherlands}

\title{Magnetoresistance in multilayer fullerene spin valves: a first-principles study}

\begin{abstract}
Carbon-based molecular semiconductors are explored for application in spintronics because their small spin-orbit coupling promises long spin life times. We calculate the  electronic transport from first principles through spin valves comprising bi- and tri-layers of the fullerene molecules C$_{60}$ and C$_{70}$, sandwiched between two Fe electrodes. The spin polarization of the current, and the magnetoresistance depend sensitively on the interactions at the interfaces between the molecules and the metal surfaces. They are much less affected by the thickness of the molecular layers. A high current polarization ($\mathrm{CP}>90$\%) and magnetoresistance ($\mathrm{MR}> 100$\%) at small bias can be attained using C$_{70}$ layers. In contrast, the current polarization and the magnetoresistance at small bias are vanishingly small for C$_{60}$ layers. Exploiting a generalized Julli{\`{e}}re model we can trace the differences in spin-dependent transport between C$_{60}$ and C$_{70}$ layers to differences between the molecule-metal interface states. These states also allow one to interpret the current polarization and the magnetoresistance as a function of the applied bias voltage.
\end{abstract}

\date{\today}
\maketitle

\section{Introduction}

Spintronics focuses on information processing with charge carrier spins.\cite{Fert08} Developments in spintronics, such as giant magnetoresistance (GMR) and tunneling magnetoresistance (TMR) in metallic spin valves have revolutionized the fields of magnetic recording and storage. Novel devices are envisioned that use injection and manipulation of spin-polarized currents in semiconductors, such as the spin transistor.\cite{Zutic04} Molecular semiconductors (MSC), i.e., semiconductors comprised of organic molecules, have caught the attention because carbon-based molecules promise to have advantages over conventional semiconductors such as Si or GaAs.\cite{Tombros07,Dediu09} The relatively weak spin-orbit coupling and hyperfine interactions in such molecules lead to long spin life times, i.e., long spin relaxation and dephasing times, which would allow for robust spin operations and read-out. The use of molecules also opens up a route towards single molecule spintronics, where individual molecules are considered for electronic functions. Indeed, magnetoresistance effects have been demonstrated at the single molecule level.\cite{Schmaus11,Kawahara12,Miyamachi12,Yoshida13}

Many experimental studies deal with vertical spin valves, where molecular layers are sandwiched between two ferromagnetic metal (FM) electrodes, and are used either as a tunnel barrier, or as charge and spin transport medium. Large magnetoresistance (MR) effects have been reported in spin valves based upon layers of organic molecules such as tris(8-hydroxy-quinolinato)-aluminium (Alq$_3$),\cite{Xiong04,Santos07,Dediu08,Schoonus09,Sun10,Barraud10,Schulz11} or fullerenes such as C$_{60}$.\cite{Gobbi11,Tran12,Zhang13,Zhang14,Li14} Similar effects have been observed in zinc methyl phenalenyl (ZMP) layers sandwiched between a FM electrode and a non-magnetic electrode, where the spin valve effect has been attributed to the special characteristics of the molecule/FM interface layer.\cite{Raman13} In phenomenological models for the observed spin transport effects the electronic structure, in particular the spin-polarization, of the MSC/FM interfaces plays a pivotal role in spin injection into the MSC.\cite{Barraud10} This has prompted the suggestion that highly spin-polarized currents in spintronic devices may be obtained by exploiting such interface interactions, which has been dubbed ``spinterface science'',\cite{Sanvito10} and has motivated research into the role played by the interfaces.\cite{Cinchitti08,Chan10,Lach12,Steil13,Dediu13,Droghetti14,Shi14}

The electronic structure of metal-organic interfaces is accessible through first-principles calculations, and can in some cases be interpreted using simple models for the energy level line-up at interfaces.\cite{Giovannetti08,Rusu10,Bokdam11} Photoemission spectroscopy or scanning tunnelling microscopy, combined with first-principles calculations, enable a detailed analysis of the spin-dependent electronic properties of metal-organic interfaces. Bonding between a molecule and a ferromagnetic metal leads to spin-split (anti)bonding states and induces a spin polarization that extends onto the molecule.\cite{Atodiresei10,Brede10,Javaid10,Tran11,Kawahara12,Schwobel12,Tran13,Djeghloul13} For instance, calculations on C$_{60}|$Fe(001) interfaces yield a magnetic moment of 0.2 $\mu_B$ induced on the C$_{60}$ molecules.\cite{Tran13} For electronic transport in spintronics devices, not the overall spin-polarization is decisive however, but the spin-polarization of the states around the Fermi energy.

First-principles transport calculations might establish the connection between such molecule-metal interface states and MR effects  in molecular spin valves. Calculations have been applied to model currents through a single molecule attached to two FM metal electrodes,\cite{Rocha05,Ning08,Koleini12,Liang12,Bagrets12} as they can be realized in STM experiments, for instance, where MR effects have been demonstrated at the single molecule level.\cite{Schmaus11,Kawahara12,Miyamachi12,Yoshida13} A single molecule is however not a good starting-point for modeling transport through MSCs, as binding a molecule to two electrodes markedly changes its electronic structure. For instance, fullerene molecules attached to two Fe electrodes result in metallic conduction,
whereas fullerene multilayers give a small, tunneling conductance.\cite{Tran12} 

We have calculated the spin-dependent  transport through multilayer graphene spin valves,\cite{Karpan07,Karpan08} 
and recently have demonstrated the feasibility of such calculations on molecular spin valves.\cite{Cakir14} In this paper we expand the scope of such calculations. In particular we focus on Fe$|$fullerene$|$Fe spin valves with bi- and tri-layers of C$_{60}$ and C$_{70}$ fullerene molecules. Fullerenes are particularly interesting molecules for applications in spintronics due to the absence of hydrogen atoms that lead to spin dephasing via hyperfine interactions. The Fe(001) surface is a well-established substrate for organic spintronics allowing for a controlled growth of fullerene layers.\cite{Tran11,Tran13} We study the links between the spin-dependent transport through these fullerene multilayers, and the electronic structure of the metal-organic interfaces by first-principles transport calculations. A generalized Julli{\`{e}}re or factorization model, defined in Sec.~\ref{sec:theory}, serves to rationalize these links, in particular when a single molecular state dominates the transport. 

The set-up of the transport calculations is discussed in Sec.~\ref{sec:compdetails}, and the results are discussed in Sec.~\ref{sec:results}. Somewhat surprisingly, there is a qualitative diference between the spin transport through C$_{60}$ and C$_{70}$ layers, which can be traced to a difference in the molecule-metal interface states. In particular, adsorption of C$_{70}$ leads to a spin-polarized interface state very close to the Fermi level that gives rise to a large current polarization (CP) and MR. In contrast, the corresponding interface state associated with adsorption of C$_{60}$ lies further from the Fermi level. That state is accessible by increasing the bias voltage over the spin valve, which however only leads to a relatively moderate CP and MR. Sec.~\ref{sec:summary} summarizes the main conclusions.

\section{Theory}
\label{sec:theory}

Following Landauer, the current through a quantum conductor $I^\sigma$ at finite bias $V$ and zero temperature, carried by independent particles with spin $\sigma=\uparrow,\downarrow$, is given by\cite{Datta95}
\begin{equation}
I^\sigma=\frac{e}{h}\sum_{\sigma}\int_{E_{F}-\frac{1}{2}eV}^{E_{F}+\frac{1}{2}eV}T^{\sigma}(E,V)dE,\label{eq:1}
\end{equation}
with $T^{\sigma}$  the transmission probability  of an electron with spin $\sigma$.  Expressed in non-equilibrium Green's functions (NEGF)\cite{Caroli71,Transiesta02}
\begin{equation}
T^\sigma=\mathrm{Tr}\left[\boldsymbol{\Gamma}^\sigma_{R}\mathbf{G}_{RL}^{\sigma,r}\boldsymbol{\Gamma}^\sigma_{L}\mathbf{G}_{LR}^{\sigma,a}\right],\label{eq:2}
\end{equation}
where $\mathbf{G}_{RL}^{\sigma,r}$ is the block of the retarded Green's function matrix connecting the
right and left leads through the quantum conductor, $\boldsymbol{G}_{LR}^{\sigma,a}=\left(\mathbf{G}_{RL}^{\sigma,r}\right)^{\dagger}$
is the corresponding advanced Green's function matrix block, and $\mathbf{\boldsymbol{\Gamma}}^\sigma_{R(L)}=-2\mathrm{Im}\boldsymbol{\Sigma}^\sigma_{R(L)}$,
with $\boldsymbol{\Sigma}^\sigma_{R(L)}$ the self-energy matrix connecting
the quantum conductor to the ideal right (left) lead.\cite{Datta95,Transiesta02,Khomyakov05}

One can rewrite this expression by formally partitioning the system in to a right and a left part and a coupling between the parts. A natural partitioning for organic spin valves is a left and a right interface, each consisting of a molecular layer adsorbed on a metal surface.\cite{Cakir14} Any molecular layers between the two interfaces are then incorporated in the coupling Hamiltonian.

In the tunneling regime, where the effects of multiple reflections between left and right parts can be neglected, it is  possible to simplify the transmission to 
\begin{equation}
T^\sigma=\mathrm{4\pi^{2}}\sum_{i,j} n_{Ri}^\sigma n_{Lj}^\sigma \left|H_{Ri,Lj}^\sigma\right|^2, \label{eq:8_1}
\end{equation}
see Appendix \ref{app:partitioning}. Here $n^\sigma_{Ri}$ and $n^\sigma_{Lj}$ are the spectral densities corresponding to states $i$ and $j$ of the right and left part, respectively.\cite{Datta95,Mahan90} The matrices $\mathbf{H}^\sigma_{RL}=\left(\mathbf{H}^\sigma_{LR}\right)^{\dagger}$ represent the coupling between the right and left parts.

Equation (\ref{eq:8_1}) can be used as a starting point to derive a generalized Julli{\`{e}}re expression for the magnetoresistance of an organic spin valve.
If the magnetization of the right electrode is reversed when switching from parallel (P) to anti-parallel (AP) configuration, it is reasonable to expect that the spectral densities of majority and minority spins are interchanged, but not altered, 
\begin{equation}
(n_{Ri}^\sigma)_{AP}\approx(n_{Ri}^{-\sigma})_{P}. \label{eq:8_1a}
\end{equation}
Following a simple tight-binding argument, the coupling matrix elements in the tunneling regime scale with the overlap between the wave functions of the left and right parts, which roughly scales as the product of these functions. If this is the case, then a decent approximation for the coupling matrix elements in the AP case should be
\begin{equation}
\left|H_{Ri,Lj}^\sigma\right|^2_{AP} \approx \left|H_{Ri,Lj}^\sigma H_{Ri,Lj}^{-\sigma}\right|_{P}. \label{eq:8_2}
\end{equation}
Using Eqs.~(\ref{eq:8_1a}) and (\ref{eq:8_2}) one can express the normalized difference $\Delta_{P/AP}=(T_P-T_{AP})/(T_P+T_{AP})$ between the transmissions $T=\sum_\sigma T^\sigma$ in the P and AP cases as
\begin{equation}
\Delta_{P/AP} =\frac{\sum_{i,j}\left( \nu_{Ri,Lj}^\uparrow - \nu_{Ri,Lj}^\downarrow \right)\left( \nu_{Lj,Ri}^\uparrow - \nu_{Lj,Ri}^\downarrow \right)}{\sum_{i,j}\left( \nu_{Ri,Lj}^\uparrow + \nu_{Ri,Lj}^\downarrow \right)\left( \nu_{Lj,Ri}^\uparrow + \nu_{Lj,Ri}^\downarrow \right)},\label{eq:8_3}
\end{equation}
in terms of the weighted densities
\begin{equation}
\nu_{Ri,Lj}^\sigma =  n_{Ri}^\sigma \left|H_{Ri,Lj}^\sigma \right|; \; \nu_{Lj,Ri}^\sigma =  n_{Lj}^\sigma \left|H_{Ri,Lj}^\sigma \right|, \label{eq:8_4}
\end{equation}
which are calculated for the P case. Equation (\ref{eq:8_3}) has the form of a generalized Julli\`{e}re expression in terms of weighted spin-polarization densities.\cite{Julliere75}  

\begin{figure*}[tb!]
\includegraphics[width=15cm]{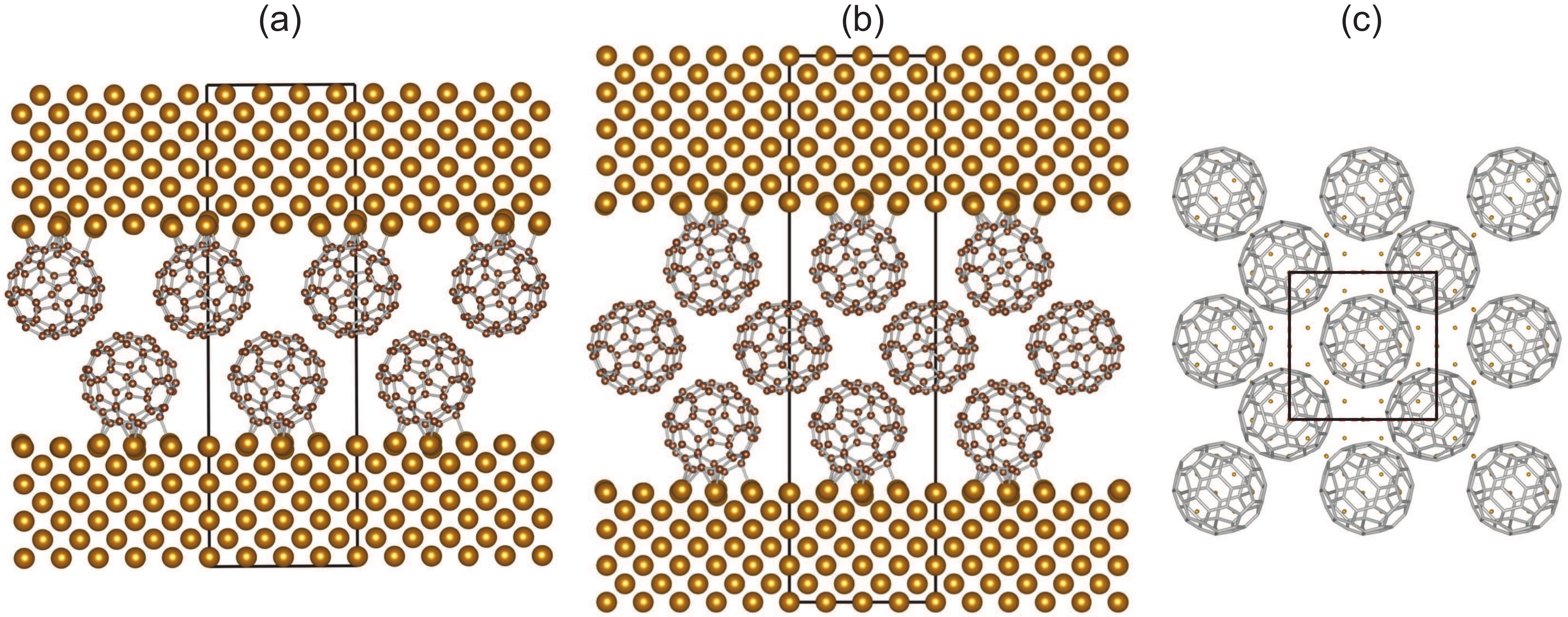}
\caption{(color online) (a),(b) side views of the bilayer and trilayer C$_{60}$ junctions, Fe$|($C$_{60})_n|$Fe, $n=2,3$; (c) top view of interlayer C$_{60}$ stacking. The black lines denote the supercell used.}
\label{fig:structures2}
\end{figure*}

The expression can be simplified if the spectral density of each spin is dominated by a single state. For adsorbed molecules this is likely to be the case for an energy range close to one particular molecular level, the HOMO or LUMO, for instance. The sum $\sum_{i,j}$ in Eq. (\ref{eq:8_3}) is then over one state, giving $\Delta_{P/AP} =P_R P_L$ with $P_{R} = (\nu^\uparrow_{R,L} - \nu^\downarrow_{R,L})/(\nu^\uparrow_{R,L} + \nu^\downarrow_{R,L})$ the weighted spin-polarization density of the right interface, where $\nu^\sigma_{R,L}=n^\sigma_{R}|H^\sigma_{RL}|$, and $P_L$ a similar expression for the spin-polarization density of the left interface. In linear response, where the bias $V$ in Eq. (\ref{eq:1}) is infinitesimal, only the transmissions at the Fermi energy are important. The magnetoresistance then becomes $\mathrm{MR}=(T_{AP}-T_{P})/T_{AP}=2P_LP_R/(1-P_LP_R)$, which has the appearance of a Julli\`{e}re expression.\cite{Julliere75} 

Assuming that a single state is dominant also allows for simplifying the transmission of  Eq. (\ref{eq:8_1}) to $T^\sigma=\mathrm{4\pi^{2}} n_{R}^\sigma n_{L}^\sigma \left|H_{RL}^\sigma\right|^2$. For a symmetric junction in the P configuration at zero bias, one has $n_L^\sigma=n_R^\sigma$, and thus $\sqrt{T_P^{\sigma}}=n_R^\sigma |H^\sigma_{RL}|$, linking the transmission directly to the interface density of states $n_R$. Applying a bias voltage $V$ across a tunnel barrier, it is reasonable to assume that the small transmission current does not change the charge distribution. The spectral densities of the right and left interfaces can then be obtained from rigid shifts of the corresponding densities at zero bias, $n^\sigma_{Ri}(E,V) = n^\sigma_{Ri}(E-eV/2,0)$ and $n^\sigma_{Lj}(E,V) = n^\sigma_{Lj}(E+eV/2,0)$. Again assuming that  at each energy a single state is dominant (not necessarily the same state at all energies), it then follows
\begin{equation}
T_P^{\sigma}\left(E,V\right) \approx \sqrt{T_P^{\sigma}\left(E-\frac{eV}{2},0\right)}\sqrt{T_P^{\sigma}\left(E+\frac{eV}{2},0\right)},\label{eq:11}
\end{equation}
and 
\begin{equation}
T_{AP}^{\sigma}\left(E,V\right)\approx\sqrt{T_P^{\sigma}\left(E-\frac{eV}{2},0\right)}\sqrt{T_P^{-\sigma}\left(E+\frac{eV}{2},0\right)}.\label{eq:10}
\end{equation}
With these expressions one can interpret the transmission spectra at any bias, starting from the spectrum of the P case at zero bias.

\section{Computational Details}
\label{sec:compdetails}

We optimize the structures of the Fe(001)$|$fullerene interfaces within density functional theory (DFT), using projector augmented waves (PAW),\cite{paw-1,paw-2} as implemented in the Vienna Ab initio Simulation Package (VASP).\cite{vasp-1,vasp-2} All plane waves up to a kinetic energy cutoff of 400 eV are included in the basis set. The spin-polarized PBE functional is used to describe exchange and correlation.\cite{pbe} As the bonding between the Fe surface and the fullerene molecules is strong, including van der Waals interactions is not necessary. An equidistant $k$-point grid with a spacing of 0.02 \AA$^{-1}$ is used for the Brillouin zone sampling. Structures are assumed to be relaxed when the difference of the total energies between two  consecutive ionic steps is less than 10$^{-5}$ eV and the maximum force on each atom is less than 0.01 eV/\AA. 

Electronic transport in Fe$|$fullerene$|$Fe junctions is calculated using the self-consistent NEGF technique, Eqs. (\ref{eq:1}) and (\ref{eq:2}), as implemented in TranSIESTA.\cite{siesta,Transiesta02} We employ Troullier-Martins (TM) normconserving pseudo-potentials (NCPP),\cite{tm} the PBE functional, and an energy cutoff for the real-space mesh of 200 Ry. Numerical orbital basis sets are used, comprising single-$\zeta$ and double-$\zeta$ plus polarization for Fe and C, respectively. To compare the VASP and SIESTA results, we benchmark the calculations on the magnetic properties of bulk bcc Fe and the clean Fe(001) surface, see Appendix \ref{app:interfaces}.

The Fe(001)$|$fullerene interfaces are modeled by a $4\times4$ Fe(001) surface unit cell, with a cell parameter of 11.32 \AA, containing one fullerene molecule, see Fig. \ref{fig:structures2}. For comparison, the nearest neighbor distance in C$_{60}$ and C$_{70}$ crystals is 10-11 \AA.\cite{Heiney91,Smaalen94}  From a number of possible adsorption geometries, we have identified the most stable structures of adsorbed C$_{60}$ and C$_{70}$ molecules. Details can be found in Appendix \ref{app:interfaces}.

A structure for a bilayer-C$_{60}$ junction, Fe$|($C$_{60})_2|$Fe, is generated by mirroring the optimized Fe(001)$|$C$_{60}$ interface structure, and translating it in plane by half a lattice constant, such that the packing C$_{60}$ molecules in the bilayer resembles that of the (001) orientation of the fcc C$_{60}$ crystal, see Fig.~\ref{fig:structures2}. The spacing between the C$_{60}$ layers is chosen such that the shortest intermolecular C--C distance is 3.2 \AA, which is a typical value for close-packed fullerenes or carbon nanotubes. Along the same lines a structure for a trilayer C$_{60}$ junction, Fe$|($C$_{60})_3|$Fe, is generated, as well as structures for bi- and trilayer C$_{70}$ junctions, Fe$|($C$_{70})_n|$Fe, $n=2,3$. 

\begin{figure}[tb!]
\includegraphics[width=7.5cm]{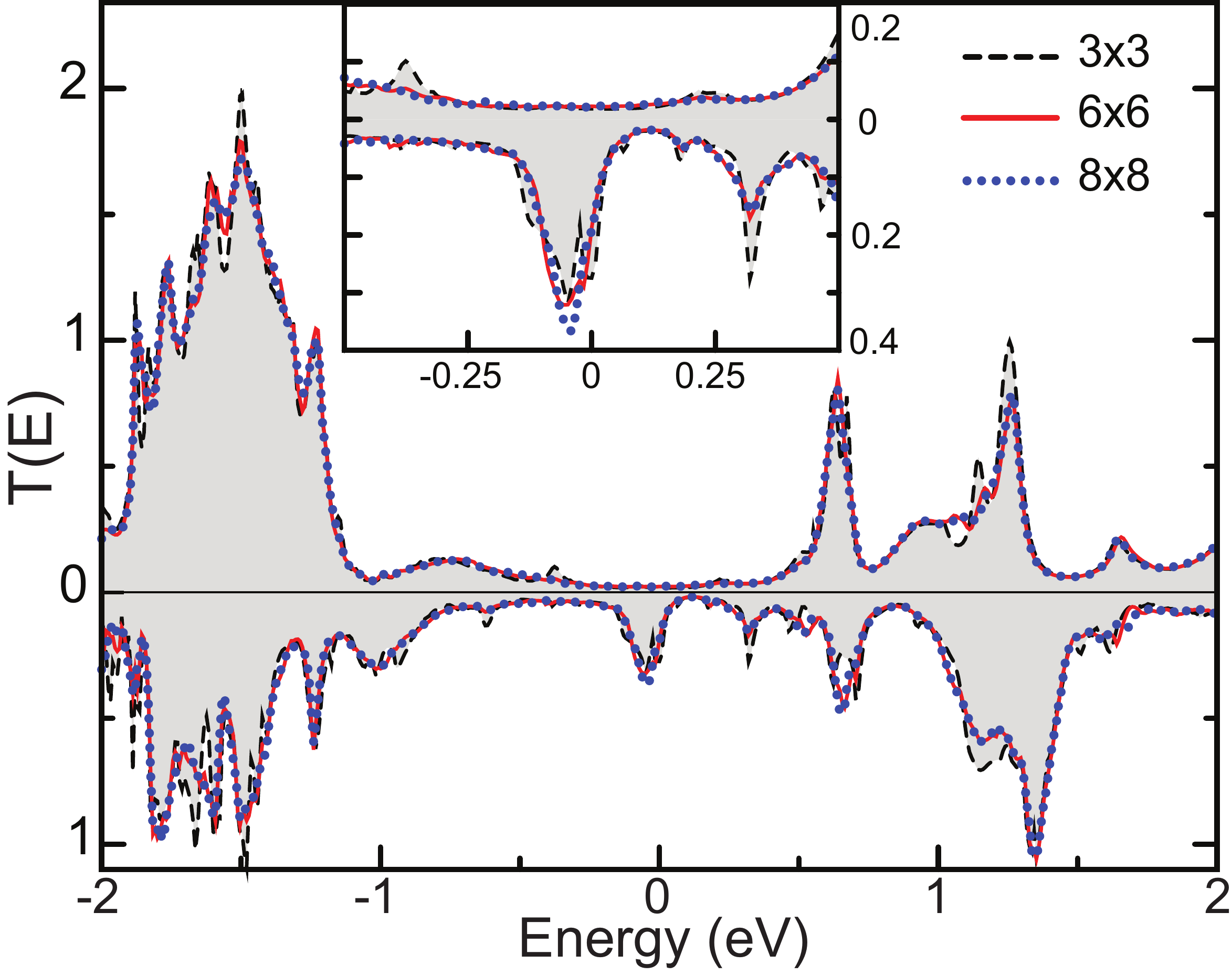}
\caption{(color online) Transmissions $T^\uparrow_P(E)$ of majority (top) and $T^\downarrow_P(E)$ (bottom) of minority spin channels of Fe$|$(C$_{70})_2|$Fe at zero bias. The Fermi level is at zero energy. Curves are given for $3\times 3$, $6\times 6$, and $8\times 8$ $k$-point grid samplings. }
\label{fig:kpointconvergence}
\end{figure}

Using a 6$\times$6 in-plane $k$-point mesh in the $4\times4$ Fe(001) supercell suffices to obtain converged results for the transmission, as is demonstrated by Fig.~\ref{fig:kpointconvergence}. 

\section{Results and discussion}
\label{sec:results}
\subsection{Linear Response}

The conductance in the linear response regime is determined by the transmission at the Fermi level. Table \ref{tab:transmission} gives the calculated transmissions of bilayers and trilayers of C$_{60}$ and C$_{70}$ molecules, sandwiched between two Fe(001) electrodes, with magnetizations parallel (P), or anti-parallel (AP). The transmission through a trilayer is up to two orders of magnitude smaller than the transmission through a bilayer. In absolute numbers the transmission through a bilayer is fairly high; the small numbers obtained for a trilayer are typical for the tunneling regime. The sizeable difference in the transmissions of bi- and tri-layers shows that we are not in the regime of resonant transmission though a molecular level. 

The transmissions of C$_{70}$ bi- or tri-layer are consistently higher than that of their C$_{60}$ counterparts, although the relevant energy levels and wave functions (HOMO and LUMO) of the isolated C$_{60}$ and C$_{70}$ molecules are not so different. Below we will argue that the difference in transmission is caused by differences in the states formed at the Fe/molecule interfaces.

\begin{table}[t]
\caption{Transmissions $T^{\uparrow(\downarrow)}_P$ of majority (minority) spins through Fe$|$layer$|$Fe at zero bias, magnetizations of electrodes parallel; transmission $T_{AP}^{\uparrow(\downarrow)}$, magnetizations anti-parallel; current polarization CP $=(T^\uparrow_P-T^\downarrow_P)/(T^\uparrow_P+T^\downarrow_P)$, normalized magnetoresistance $\Delta_{P/AP}=(T_P-T_{AP})/(T_P+T_{AP})$, and optimistic magnetoresistance MR $=(T_P-T_{AP})/T_{AP}$ (in \%).  }
\begin{ruledtabular}
\begin{tabular}{lccccccc}
layer        &   $T_P^\uparrow(E_F)$ & $T_P^\downarrow$ &   $T_{AP}^{\uparrow(\downarrow)}$ & $\sqrt{T_P^\uparrow T_P^\downarrow}$ & CP & $\Delta_{P/AP}$ & MR \\
\hline                                                                                                                          
(C$_{60}$)$_2$        &   $8.6(-3)$\footnotemark[1]  &  $8.4(-3)$  &   $9.1(-3)$    &   $8.5(-3)$  & 1.3         & $-3.6$ & $-6.9$\\
(C$_{60}$)$_3$        &   $2.1(-4)$  &  $2.9(-4)$  &   $2.5(-4)$    &   $2.5(-4)$  & $-16$    & $1.1$   & $2.2$\\
(C$_{70}$)$_2$        &   $2.3(-2)$  &  $1.9(-1)$  &   $6.3(-2)$    &   $6.6(-2)$  & $-78$    & $25$   & $67$\\
(C$_{70}$)$_3$        &   $2.9(-4)$  &  $4.1(-3)$  &   $9.6(-4)$    &   $11.0(-4)$  & $-87$  & $42$   & $144$\\     
\end{tabular}
\end{ruledtabular}
\footnotetext[1]{$8.6\times 10^{-3}$}
\label{tab:transmission}
\end{table}

The most prominent difference between C$_{60}$ and C$_{70}$ molecules is in the spin polarization of the transmission. Whereas for C$_{60}$ layers the transmissions of majority and minority spins are almost equal, for C$_{70}$ layers the transmission of minority spin is approximately an order of magnitude larger than that of majority spin. It means that the current polarization in linear response, $\mathrm{CP}=(T^\uparrow_P-T^\downarrow_P)/(T^\uparrow_P+T^\downarrow_P)$, of C$_{60}$ junctions at low bias is small, $|\mathrm{CP}| < 20 $\%. In contrast, the CP of C$_{70}$ junctions is very substantial, $|\mathrm{CP}| = 80$-$90$\%. Moreover the magnetoresistance $\mathrm{MR}=(T_P-T_{AP})/T_{AP}$ is large for C$_{70}$ junctions, exceeding 100\% for trilayers, whereas the MR for C$_{60}$ junctions is vanishingly small. The differences between the CP and MR of C$_{60}$ and C$_{70}$ junctions have the same origin, as we will see below.

\begin{figure}[t]
\includegraphics[width=8.5cm]{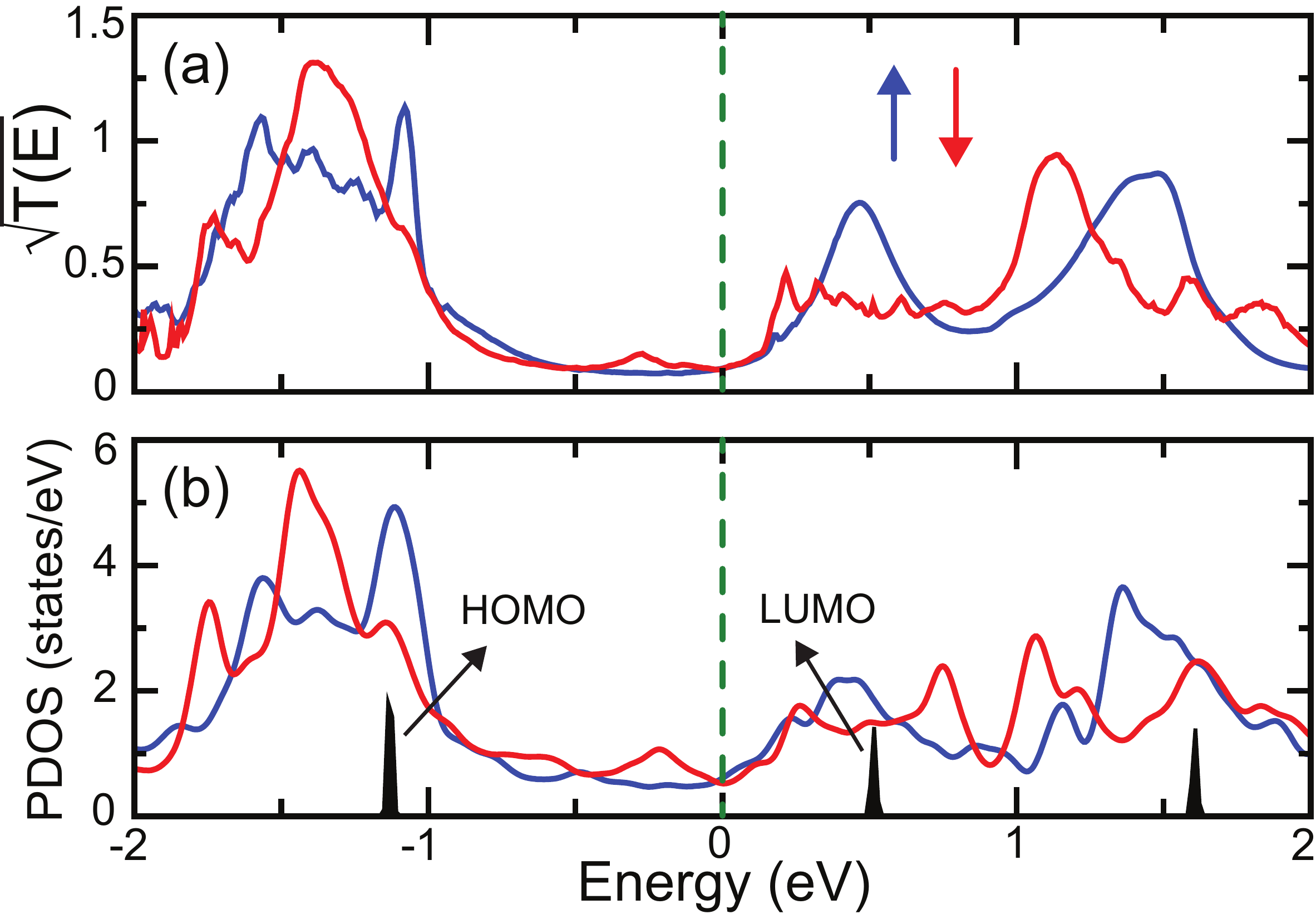}
\caption{(color online) (a) $\sqrt{T_P^\uparrow(E)}$ for majority spin (blue) and $\sqrt{T_P^\downarrow(E)}$ for minority spin (red) at zero bias for Fe$|$C$_{60}$-C$_{60}|$Fe junction; (b) projected density of states (PDOS) of the Fe$|$C$_{60}$ interface.}
\label{fig:squareT}
\end{figure}

In the following we interpret the behavior of C$_{60}$ and C$_{70}$ junctions using the model outlined in Sec. \ref{sec:theory}. If a single channel dominates the transmission and the junction is symmetric, then transmission can be factorized according to Eqs.~(\ref{eq:11}) and (\ref{eq:10}), and the factors $\sqrt{T_P^{\sigma}}=n_R^\sigma |H^\sigma_{RL}|$, are weighted interface density of states. Figure~\ref{fig:squareT}(a) shows $\sqrt{T_P^{\sigma}(E)}$, derived from the transmission spectra of a bilayer C$_{60}$ junction. For comparison, Fig.~\ref{fig:squareT}(b) shows the density of states of a C$_{60}$/Fe(001) interface, projected on the molecule (PDOS), see Appendix \ref{app:interfaces}. There is indeed a striking resemblance between the transmission spectra and the PDOSs. 

The peaks in the PDOS can be labeled according to their molecular character. As the molecule-substrate interaction is large, these peaks correspond to hybrid interface states, which are significantly broadened in energy compared to the pure molecular states. Moreover, the interface states are exchange-split because the substrate is ferromagnetic. Nevertheless the dominant components of their molecular contributions can still be identified; details can be found in Appendix \ref{app:interfaces}. 

The Fermi level is situated in a gap in the transmission spectra of the bilayer C$_{60}$ junction, which according to the PDOS and the molecular level spectrum corresponds to the HOMO-LUMO gap. One has $T_P^\uparrow(E_F) \approx T_P^\downarrow(E_F)$, and this absence of spin polarization is also observed in the PDOS. It is then not surprising to find that $\mathrm{CP}\approx 0$ and $\mathrm{MR}\approx 0$ at low bias. The two small peaks in the minority spin transmission $T_P^\downarrow(E)$ at $E\approx E_F\pm 0.2$ eV will give rise to a moderate nonzero CP and MR at finite bias, as we will see in the next section. These peaks are derived from hybridyzing the Fe(001) surface states with the C$_{60}$ LUMO. 

\begin{figure}[t]
\includegraphics[width=8.0cm]{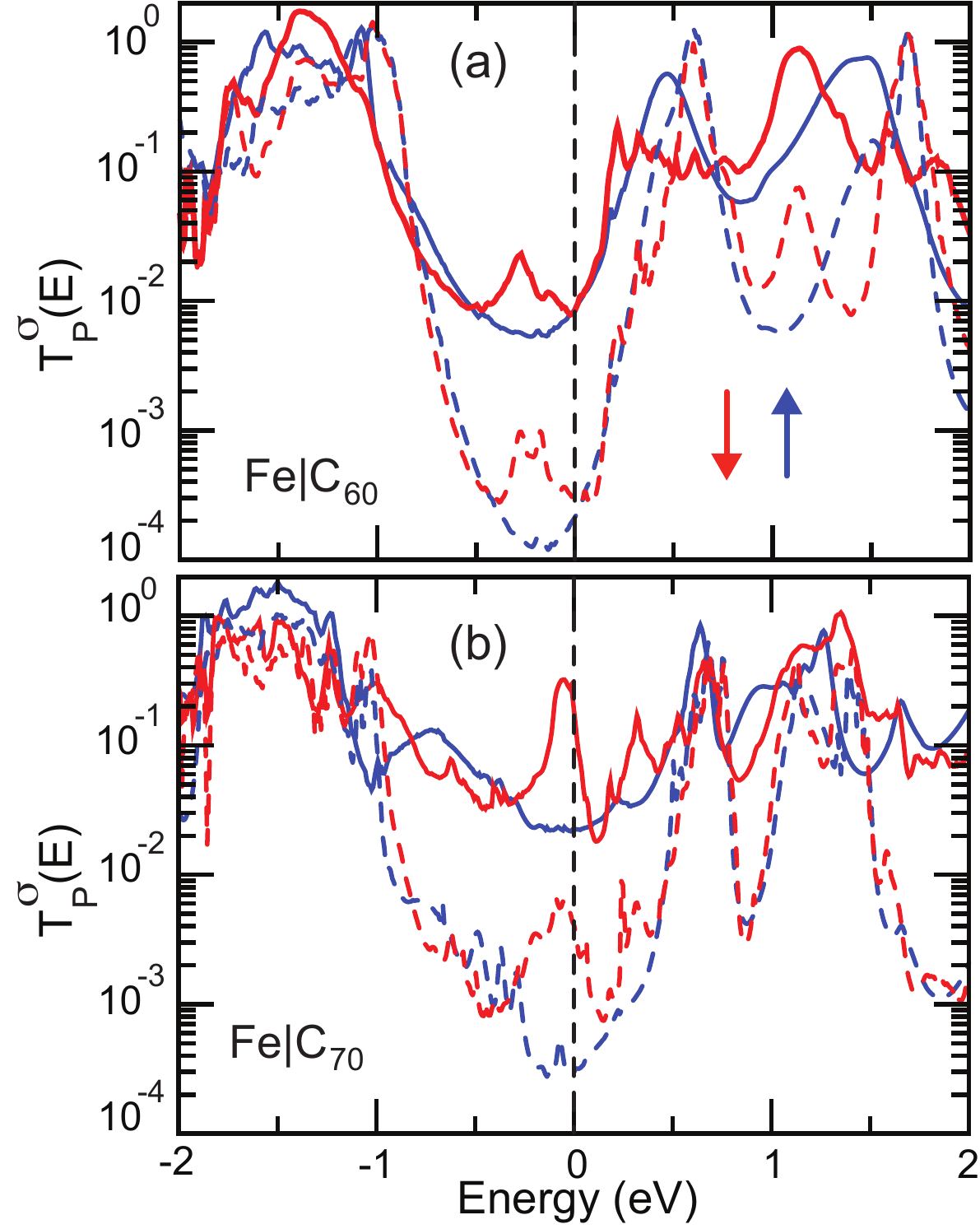}
\caption{(color online) (a) Calculated transmission spectra $T_P^\sigma(E)$ of majority (blue) and minority (red) spins of bilayer C$_{60}$ (solid lines) and trilayer C$_{60}$ (dashed lines) junctions. The Fermi level $E_F$ is at zero energy. (b) Calculated transmission spectra of C$_{70}$ junctions.}
\label{fig:transmission}
\end{figure}

Figure~\ref{fig:transmission} gives the transmission spectra $T_P^\sigma(E)$ of all the multilayers studied in this paper. The peaks in the transmission spectra of order unity correlate with resonant transmission through molecular levels. In the bilayer case the latter are strongly hybridized with the Fe surface, resulting in broad peaks. The transmission spectra for bi- and tri-layers are qualitatively similar, but for the trilayer the peaks in the transmission are considerably sharper. For the trilayer transmission of order unity can only be achieved via resonant transmission through the molecular levels of the middle layer. 

The transmission for energies in the gaps between the peaks imply tunneling through the molecular layers. In all cases the Fermi level is situated in the gap in the transmission spectrum corresponding to the molecular HOMO-LUMO gap. The transmission for energies inside this gap is higher for C$_{70}$ layers than for C$_{60}$ layers. This is consistent with the difference between these molecules regarding the spatial extent of their interface states. The interaction between C$_{70}$ and the Fe(001) surface gives interface states that are more delocalized over the molecules, see Appendix \ref{app:interfaces}. Such a delocalization effectively leads to thinner tunnel barriers. 

\begin{figure}[t]
\includegraphics[width=8.0cm]{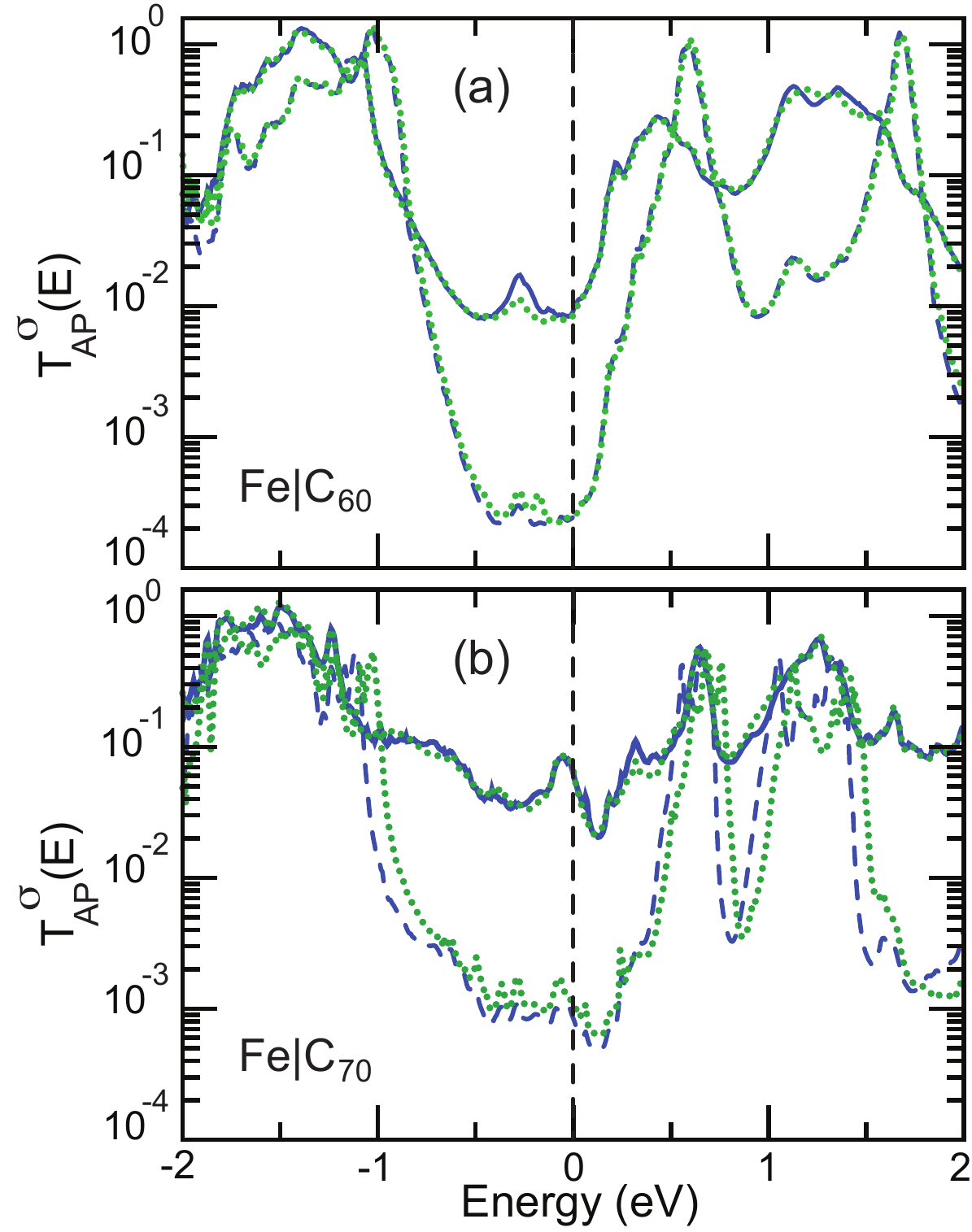}
\caption{(color online) Calculated transmission spectra $T_{AP}^\sigma(E)$ (blue) of bilayer (solid lines) and trilayer (dashed lines) molecular junctions of (a) C$_{60}$ and (b) C$_{70}$.\cite{footnoteAP} The Fermi level $E_F$ is at zero energy. The green dotted lines give the factorization approximation $2\sqrt{T_{P}^\sigma(E)T_{P}^{-\sigma}(E)}$, see Eq. (\ref{eq:10}).}
\label{fig:APtransmission}
\end{figure} 

The most prominent difference between C$_{60}$ and C$_{70}$ in the transmission close to the Fermi level is a peak in the minority spin channel, compare Figs.~\ref{fig:transmission}(a) and (b). For C$_{70}$ a prominent peak in the transmission is situated very close to the Fermi level, whereas for C$_{60}$ a smaller peak lies at $\sim 0.2$ eV below the Fermi level. Both these peaks can be traced to an interface state derived from the molecular LUMO, created by the adsorption of the molecules on the surface. Differences in the bonding of the two molecules to the Fe(001) surface give a different energy for this state, which has a major effect on the spin transport properties of the molecular layers, see Table \ref{tab:transmission}. For C$_{70}$ this minority spin state at the Fermi level is at the origin of a large CP and a large MR, whereas for C$_{60}$ the fact that this state is not exactly at the Fermi level results in a small CP and a small MR. Going from two to three layers the transmission in the HOMO-LUMO gap decreases, but the overall pattern of the transmission remains the same.  

Figure \ref{fig:APtransmission} shows the transmission spectra $T_{AP}^\sigma(E)$ calculated with the magnetizations of the two Fe electrodes in anti-parallel configurations. Also shown are the results of the factorization approximation, Eq. (\ref{eq:10}), with $V=0$. Following the discussion in Sec. \ref{sec:theory} this approximation is designed for the tunneling regime, when multiple reflections are absent, and when a single channel dominates the transmission. The results shown in Fig. \ref{fig:APtransmission} seem to indicate that the factorization approximation has a somewhat wider applicability, and also works reasonably well if the transmission is larger than is typical for tunneling. From the  factorization approximation it becomes clear that the CP and the MR are related properties. If the CP is large (small), then the MR is large (small).

There are of course situations where the factorization approximation fails. For instance it always give a $\mathrm{MR}\geq 0$ for a symmetric junction in the linear response regime. This is easy to see from the discussion following Eqs.~(\ref{eq:8_3}) and (\ref{eq:8_4}). In a symmetric junction at zero bias, the weighted spin-polarizations of left and right interfaces is identical, $P_R=P_L$, which implies that $\Delta_{P/AP}\geq 0$ and $\mathrm{MR}\geq 0$. The small negative MR at zero bias calculated for a bilayer C$_{60}$ junction in Table \ref{tab:transmission} is clearly in disagreement with this. By construction this junction is symmetric, and the right  Fe(001)$|$C$_{60}$ interface is identical to the left interface.

Nevertheless it is possible to obtain a negative MR, even for a symmetric junction. To obtain $\Delta_{P/AP} < 0$ in Eq.~(\ref{eq:8_3}), giving $\mathrm{MR}< 0$, (at least) two molecular states at each interface should be involved. In absence of any off-diagonal coupling, i.e., $H_{Ri,Lj}^\sigma=0$; $i\neq j$, this would still give $\Delta_{P/AP}\geq 0$, so in order to obtain a negative sign one needs a significant off-diagonal coupling. Suppose for simplicity that $H_{Ri,Lj}^\sigma=h$; $i\neq j$, and $H_{Ri,Li}^\sigma=0$, then in a system with two states it suffices to have $n_1^\downarrow > n_1^\uparrow$ and $n_2^\downarrow < n_2^\uparrow$ to obtain $\Delta_{P/AP} < 0$. In other words, a negative MR in a symmetric junction can be obtained if a strong coupling between two states exists, where one of the states has a dominant majority spin character at the Fermi level, and the other one has a dominant minority spin character.

As in the bilayer C$_{60}$ junction the Fermi level falls between peaks in the transmission spectra and in the PDOS, see Fig.~\ref{fig:squareT}, it is quite likely that the tails of more than one molecular state are involved at this energy. The fact that in the trilayer C$_{60}$ junction the negative MR disappears shows that the coupling between these states across the junction is crucial. Despite its limitations, the factorization model can be very helpful in interpreting spin transport properties as we will see in the next section. 

\subsection{Finite bias}

\begin{figure}[t]
\includegraphics[width=6.0cm]{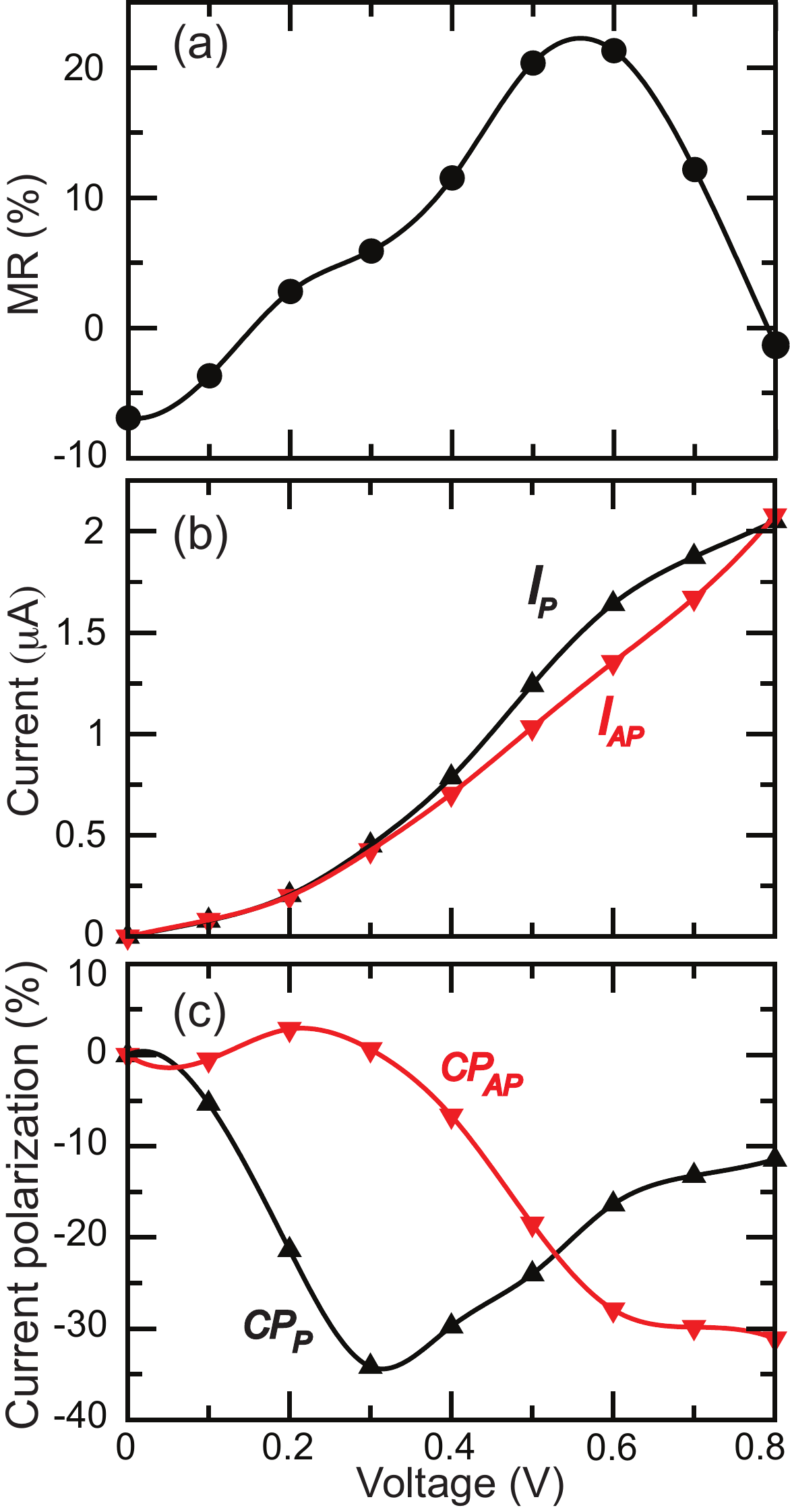}
\caption{(color online) (a) Magneto-resistance $\mathrm{MR} = (I_P-I_{AP})/I_{AP}$ of the C$_{60}$ bilayer junction as function of bias $V$. (b) Total currents $I_P$ (black) and $I_{AP}$ (red)  for the magnetizations of both electrodes parallel, respectively anti-parallel. (c) Current polarization $\mathrm{CP} = (I^\uparrow-I^\downarrow)/(I^\uparrow+I^\downarrow)$.}
\label{fig:mr}
\end{figure}
 
Figure \ref{fig:mr}(a) shows the magneto-resistance (MR) as a function of the applied bias $V$, calculated self-consistently for the Fe$|$bilayer C$_{60}|$Fe junction. Over the voltage range studied the MR increases from $-7$\% at $V=0$ to $+21$\% at $V=0.6$ V, before dropping again to $-1$\% at $V=0.8$ V. The calculated total currents $I_P$ and $I_{AP}$ as a function of the applied bias $V$ are shown in Fig. \ref{fig:mr}(b). The currents are distinctly non-linear, and the junction is non-ohmic. The transmissions of the bilayer C$_{60}$ junction at energies in the interval $E_F\pm 0.5$ eV are of order $10^{-2}$, see Figs. \ref{fig:transmission}(a) and \ref{fig:APtransmission}(a), which, although much smaller than unity, is still larger than is typical for a tunnel junction. In other words, a bilayer C$_{60}$ junction is still quite a leaky junction. 

Figure \ref{fig:mr}(c) shows the spin polarization of the current or current polarization (CP). At zero bias, $V=0$, the total current $I_P=I^\uparrow+I^\downarrow$ through a parallel junction is not polarized, i.e., $I^\uparrow=I^\downarrow$. Upon increasing the bias the minority spin-current becomes dominant, $I^\downarrow>I^\uparrow$, and the total current $I_P$ becomes polarized with a minimum of $-35$\% at $V=0.3$ V. Remarkably, a current polarization of a similar magnitude can be achieved in an AP junction, albeit at a bias that is more than twice as large.

\begin{figure}[t]
\includegraphics[width=8.5cm]{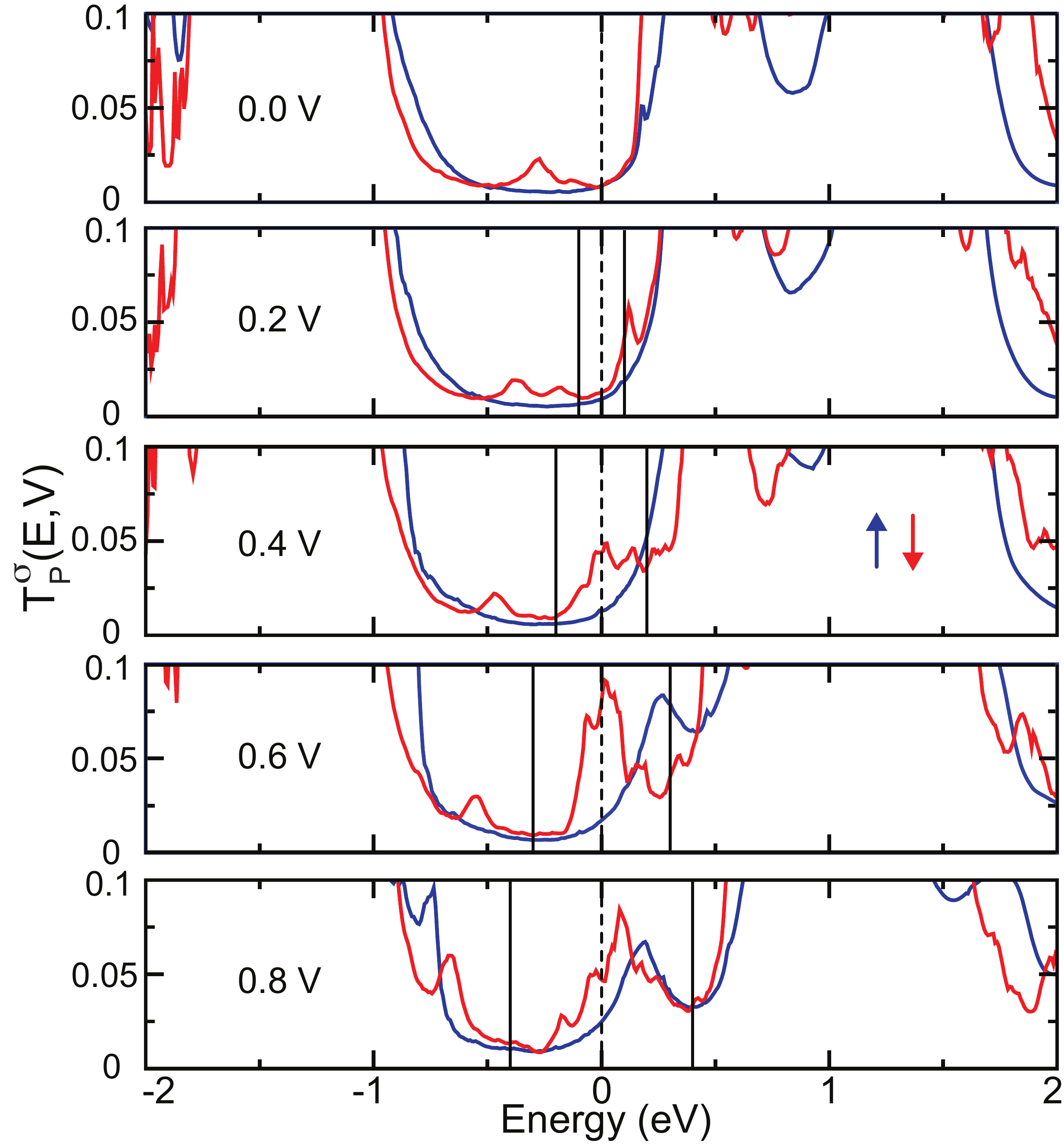}
\caption{(color online) $T_P^\uparrow(E,V)$ for majority spin (blue) and $T_P^\downarrow(E,V)$ for minority spin (red) of the C$_{60}$ bilayer junction as function of bias, from top to bottom: $V=0.0$, 0.2, 0.4, 0.6, 0.8 V. The vertical lines enclose the energy interval over which to integrate to obtain the total current according to Eq.~(\ref{eq:1}).}
\label{fig:TVdependence}
\end{figure}

Figure~\ref{fig:TVdependence} shows the transmission spectra $T_P^\sigma(E,V)$ at finite bias, calculated self-consistently, for a range of different biases. The factorization model allows one to interpret the trends in these spectra, and in the MR and the CP. According to Eq.~(\ref{eq:11}) one can construct the transmission at finite bias starting by multiplying a pair of $\sqrt{T_P^{\sigma}}$ spectra, displaced by $\pm eV/2$, cf. Fig.~\ref{fig:squareT}. At zero bias the CP is zero, reflecting the fact that $T_P^\uparrow(E_F,0)=T_P^\downarrow(E_F,0)$. Close to the Fermi level at $E\approx E_F\pm 0.2$ eV, the minority spin transmission $T_P^\downarrow(E,0)$ show two small peaks, Fig.~\ref{fig:squareT}(a). Both these peaks are derived from interface states involving the C$_{60}$ LUMO, as discussed in Appendix \ref{app:interfaces}.  

Increasing the bias means that these two peaks move towards one another, according to Eq.~(\ref{eq:11}). If one displaces the $T_P^\downarrow(E,0)$ spectra by $\pm 0.2$ eV, these two peaks coincide at the same energy. According to Eq.~(\ref{eq:11}) such a displacement corresponds to a bias of 0.4 V. As the majority spin transmission $T_P^\uparrow(E,0)$ does not have peaks in this energy region, it means that at this bias the CP is negative, which indeed is the case, as can be seen in Fig. \ref{fig:mr}(c). Upon increasing the bias further, peaks in the majority spin transmission also move into the integration window for the total current. This means that the CP decreases with increasing bias, which can also be observed Fig. \ref{fig:mr}(c).  

\begin{figure}[t]
\includegraphics[width=8.5cm]{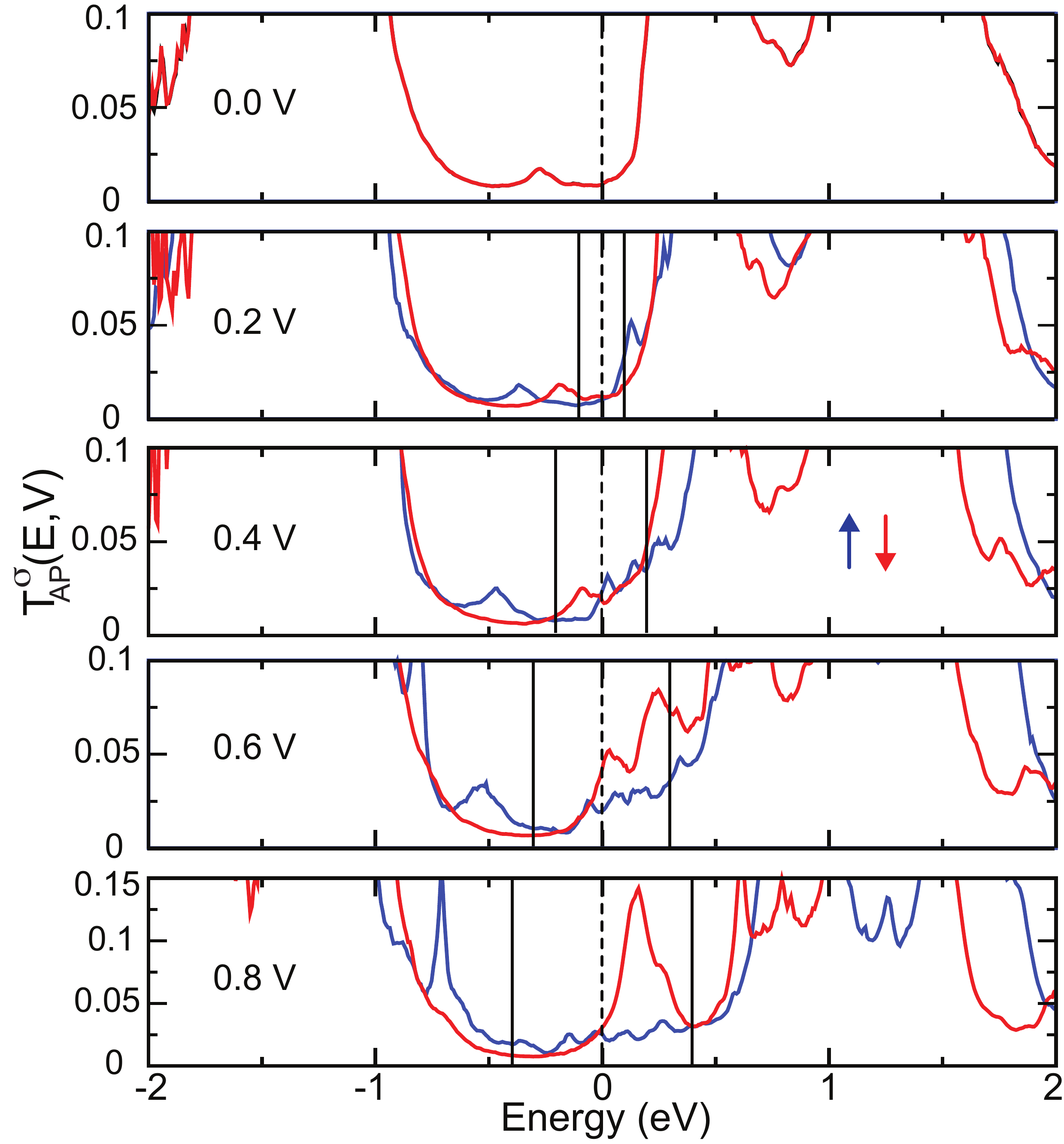}
\caption{(color online) $T_{AP}^\uparrow(E,V)$ for majority spin (blue) and $T_{AP}^\downarrow(E,V)$ for minority spin (red) of the C$_{60}$ bilayer junction as function of bias, from top to bottom: $V=0.0$, 0.2, 0.4, 0.6, 0.8 V. The vertical lines enclose the energy interval over which to integrate to obtain the total current according to Eq.~(\ref{eq:1}).}
\label{fig:TVdependenceAP}
\end{figure}

To explain the CP for the AP case one must realize that the role of majority and minority spin are now interchanged for one of the electrodes. Within the factorization model, one has to multiply a $\sqrt{T_P^\sigma}$ curve with a $\sqrt{T_P^{-\sigma}}$ curve, displaced by $\pm eV/2$, see Fig.~\ref{fig:squareT}. The first peak in $T_P^\uparrow$ is at $E\approx E_F + 0.5$ eV. In order to have that coincide with the peak in $T_P^\downarrow$ at $E\approx E_F - 0.2$~eV, it requires a bias $V\approx 0.7$ V. Multiplying the $\sqrt{T_P^\sigma}$ and $\sqrt{T_P^{-\sigma}}$ factors then gives a peak in the transmission spectrum $T_{AP}^\downarrow(E,V)$ at $E=0, V= 0.7$ V. Indeed the self-consistent transmission spectrum $T_{AP}^\sigma$ at finite bias, given in Fig.~\ref{fig:TVdependenceAP}, shows this peak in $T_{AP}^\downarrow$ growing with increasing bias. Therefore, for the AP case one expects to see a zero CP at low bias, and a decreasing CP at higher bias, which is indeed the case in Fig.~\ref{fig:mr}(c).

The behavior of the MR in Fig.~\ref{fig:mr}(a) can be interpreted qualitatively along the same lines. The MR is zero at zero bias because the transmission of both spin channels is almost the same. Upon increasing the bias foremost the transmission of the minority spin channel in the P case increases, cf. Fig.~\ref{fig:TVdependence}, which according to the factorization model originates from shifting the two minority spin peaks at $E\approx E_F\pm 0.2$~eV closer together, see Fig.~\ref{fig:squareT}(a), as discussed above. There is not such an increase in the AP case, as the roles of majority and minority spin in one of the electrodes are reversed. This means that, upon increasing the bias, $I_{AP}<I_{P}$ as can be observed in Fig.~\ref{fig:mr}(b). Upon further increase of the bias the transmission of the minority spin channel in the AP case increases, cf. Fig.~\ref{fig:TVdependenceAP}, as the minority spin peak at $E\approx E_F - 0.2$~eV starts to approach the majority spin peak at $E\approx E_F + 0.5$ eV, as discussed in the previous paragraph. It means that at a higher bias $I_{AP}$ increases relative to $I_{P}$, and the MR decreases again, see Fig.~\ref{fig:mr}(a).

The behavior of the Fe(001)$|$bilayer C$_{70}|$ junction as a function of bias voltage has been explained in Ref.~\onlinecite{Cakir14}. It is much simpler than that of the bilayer C$_{60}$ junction. At zero bias a peak in the transmission for minority spin is found very close to the Fermi level, which results in a substantial CP and MR at zero bias. Upon increasing the bias this peak decreases as in the factorization model the two factors are displaced from one another, cf. Eq.~(\ref{eq:10}). That results in a monotonic derease of both $|\mathrm{CP}|$ and $|\mathrm{MR}|$ as a function of bias.

The transmission spectra of the molecular trilayer junctions are qualitatively similar to those of their corresponding bilayers, see Figs.~\ref{fig:transmission} and \ref{fig:APtransmission}. This means that as a function of bias one expects the CP and the MR of trilayers to behave similarly to their bilayer counterparts. Of course the absolute currents for the trilayer cases will be much lower than for the bilayer cases. The fact that the general behavior of the CP and the MR does not depend critically on the thickness of the molecular layers illustrates the central role played by the metal-molecule interfaces. 

\section{Summary and Conclusions}
\label{sec:summary}
We calculate the electronic transport from first principles through spin valves composed of bilayers and trilayers of the fullerene molecules C$_{60}$ and C$_{70}$, sandwiched between two ferromagnetic Fe electrodes. Despite the similarity of the two molecules, they give rise to a markedly different behavior of their spin-dependent currents.  C$_{70}$ bi- and tri-layers give large negative current polarizations of $-80$ to $-90$\% at small bias, where the minus sign indicates that the currents are dominated by minority spin. In contrast, the current polarization generated by C$_{60}$ layers is zero at small bias. Similarly, the magnetoresistance of C$_{70}$ spin valves at small bias is $70$ to $140$\%, whereas that of C$_{60}$ spin valves is only a few percent. As a function of applied bias across the spin valve, the current polarization of C$_{70}$ junctions increases monotonically toward zero, and the magnetoresistance decreases toward zero. For bilayer C$_{60}$ spin valves the current polarization goes through a minimum of $-35$\% at $V=0.3$ V, as a function of applied bias, and the magnetoresistance goes through a maximum of $24$\% at $V=0.55$ V.

All these trends can be explained using a generalized Julli{\`{e}}re or factorization model, which couples the spin-dependent transport of the junctions to the electronic structure of the molecule-metal interfaces. The favorable properties of C$_{70}$ junctions can be traced to an interface state in the minority spin, which is derived from the molecular LUMO, and lies very close the Fermi energy. Although a similar state also exists for C$_{60}$, it lies $\sim 0.2$~eV below the Fermi level, which means that it becomes accessible only at a higher bias voltage. The binding of the molecules to the surface plays a decisive role in determining the position of these states with respect to the Fermi level. Increasing the thickness of the molecular layers decreases the abolute value of the currents, but it has a relative small effect on the sizes of the current polarization and of the magnetoresistance, which stresses the pivotal role played by the molecule-metal interfaces.

\begin{acknowledgments}
We thank Michel de Jong and Zhe Yuan for useful discussions. 
Computational resources were provided through the Physical Sciences division of the Netherlands Organization for Scientific Research (NWO-EW) and by TUBITAK ULAKBIM, High Performance and Grid Computing Center (TR-Grid e-Infrastructure). 
\end{acknowledgments}   

\appendix
\section{Partitioning}\label{app:partitioning}
To calculate the transmission, Eq.~(\ref{eq:2}), one needs the block of the Green's function matrix $\mathbf{G}_{RL}$ connecting the right and left leads via the quantum conductor, where we omit the spin index for the moment to simplify the notation. We partition the system in to a left and a right part, 
\begin{equation}
\left(\begin{array}{cc}
\mathbf{G}_{LL} & \mathbf{G}_{LR}\\
\mathbf{G}_{RL} & \mathbf{G}_{RR}\end{array}\right)
\left(\begin{array}{cc}
E\mathbf{I}_L-\mathbf{H}_{LL} & -\mathbf{H}_{LR}\\
-\mathbf{H}_{RL} & E\mathbf{I}_R-\mathbf{H}_{RR}\end{array}\right)=\left(\begin{array}{cc}
\mathbf{I}_{L} & 0\\
0 & \mathbf{I}_{R}\end{array}\right),\label{eq:3}
\end{equation}
where the diagonal blocks $\mathbf{H}_{LL}$ and $\mathbf{H}_{RR}$ of the Hamiltonian
matrix represent the semi-infinite left and right parts, and the off-diagonal blocks $\mathbf{H}_{RL}=\left(\mathbf{H}_{LR}\right)^{\dagger}$
represent the coupling between the right and left parts. 

Formally solving Eq.~(\ref{eq:3}) then gives for the off-diagonal block of the Green's function matrix  
\begin{equation}
\mathbf{G}_{RL} = \mathbf{g}_{R}\mathbf{H}_{RL}\left(\mathbf{I}_L-\mathbf{g}_{L}\mathbf{H}_{LR}\mathbf{g}_{R}\mathbf{H}_{RL}\right)^{-1}\mathbf{g}_{L},\label{eq:4} 
\end{equation}
with 
\begin{equation}
\mathbf{g}_{R} = \left(E\mathbf{I}_R-\mathbf{H}_{RR}\right)^{-1}; \; \mathbf{g}_{L} = \left(E\mathbf{I}_L-\mathbf{H}_{LL}\right)^{-1}, \label{eq:5} 
\end{equation}
the Green's function matrices of the uncoupled right and left parts. Expression (\ref{eq:4}) can be used to rewrite  Eq.~(\ref{eq:2}). Moreover it is easy to show that $\mathbf{g}_{R(L)}^{a}\mathbf{\boldsymbol{\Gamma}}_{R(L)}\mathbf{g}_{R(L)}^{r}=2\pi\mathbf{n}_{R(L)}$, where $\mathbf{n}_{R(L)}=-\pi^{-1}\mathrm{Im}\mathbf{g}_{R(L)}^{r}$ is the spectral density matrix of the right (left) part.\cite{Datta95,Mahan90} Equation (\ref{eq:2}) then becomes
\begin{eqnarray}
T=\mathrm{4\pi^{2}Tr}\biggl[ \mathbf{n}_{R}\mathbf{H}_{RL}\bigl(\mathbf{I}_L&-&\mathbf{g}_{L}^{r}\mathbf{H}_{LR}\mathbf{g}_{R}^{r}\mathbf{H}_{RL}\bigr)^{-1} \nonumber \\
 \mathbf{n}_{L}\mathbf{H}_{LR} \bigl( \mathbf{I}_R&-&\mathbf{g}_{R}^{a}\mathbf{H}_{RL}\mathbf{g}_{L}^{a}\mathbf{H}_{LR} \bigr) ^{-1} \biggr].\label{eq:6}
\end{eqnarray}
A similar expression has been derived in Ref. \onlinecite{Mingo96} to model scanning tunneling microscopy. It can also be derived from the (linear response) Kubo formalism, as in Refs. \onlinecite{Mathon97,Mathon99}. The expression is however also valid outside the linear response regime, cf. Eq. (\ref{eq:1}), provided the density, Hamiltonian, and Green's function matrices are calculated self-consistently.\cite{Transiesta02}

The terms $(\mathbf{I}_L-\cdots )^{-1}$ and $(\mathbf{I}_R-\cdots)^{-1}$  in Eq. (\ref{eq:6}) incorporate the effects of (multiple) reflections between the left and right parts. Neglecting these, i.e. replacing these terms by $\mathbf{I}_L$ and $\mathbf{I}_R$ respectively, then gives
\begin{equation}
T=\mathrm{4\pi^{2}Tr}\left[\mathbf{n}_{R}\mathbf{H}_{RL}\mathbf{n}_{L}\mathbf{H}_{LR}\right].\label{eq:8}
\end{equation}
One expects this approximation to be accurate in the tunneling regime. Reintroducing the spin index $\sigma$, and choosing representations where the density matrices are diagonal, $(\mathbf{n}^\sigma_{R})_{ij}=\delta_{ij} n^\sigma_{Ri}$, $(\mathbf{n}^\sigma_{L})_{ij}=\delta_{ij} n^\sigma_{Lj}$, then gives Eq.~(\ref{eq:8_1}).      

\section{Fullerene$|$Fe(001) interfaces}
\label{app:interfaces}

\begin{figure}[tb]
\includegraphics[width=8.0cm]{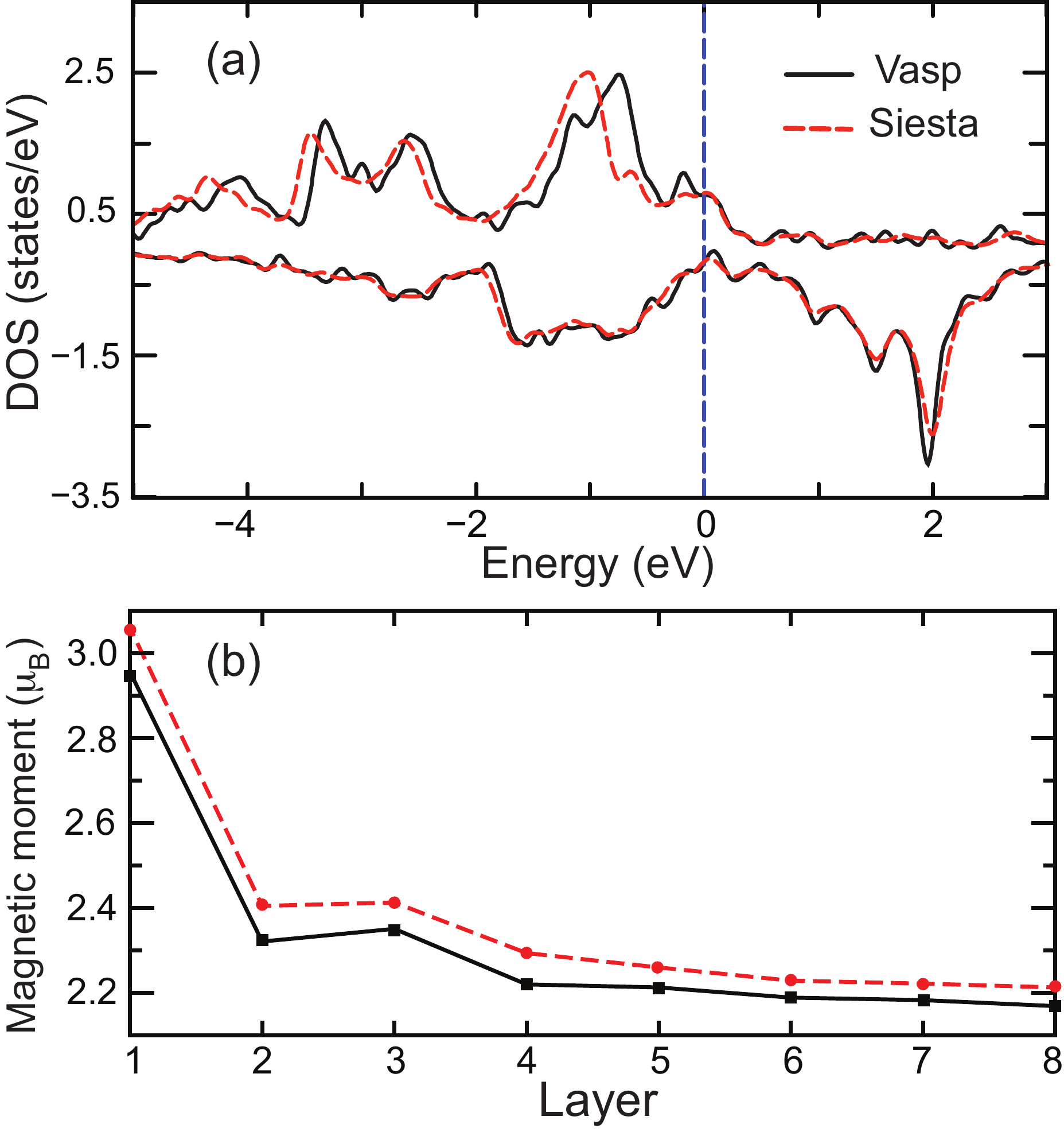}
\caption{(color online) (a) Density of states of bulk bcc Fe, calculated with PAW (black) and NCPP (red). The NCPP gives an exchange splitting that is $\sim 0.2$ eV larger, which results in a $\sim 0.05$ $\mu_B$ larger magnetic moment. (b) Magnetic moments (in $\mu_B$/atom) of the Fe(001) surface as function of layer position (1 indicates the surface layer) calculated with PAW (black) and NCPP (red).}
\label{fig:magmom}
\end{figure}

We optimize all structures with VASP,\cite{vasp-1,vasp-2} using the PBE functional and the parameter settings given in Sec.~\ref{sec:compdetails}. The optimized lattice constant of bulk Fe is $2.83$ \AA, which is in good agreement with the experimental values of 2.87 \AA.\cite{Kittel86} The magnetic moments per atom of bulk Fe are 2.20 $\mu_B$ (VASP) and  2.25 $\mu_B$ (SIESTA), respectively, which both are in good agreement with the experimental value of 2.22 $\mu_B$.\cite{Kittel86} The difference between the magnetic moments calculated with VASP and SIESTA can be traced to the use of norm-conserving pseudopotentials (NCPPs) in SIESTA, versus (all-electron) projector augmented waves (PAWs) in VASP. The former gives a larger exchange splitting, see Fig. \ref{fig:magmom}, which gives a larger magnetic moment. VASP calculations with NCPP pseudo-potentials give a similar exchange splitting as SIESTA,\cite{dewijs} so the use of different basis sets in VASP and SIESTA, i.e., plane waves versus localized atomic orbitals, is of less importance. 

The difference in calculated magnetic moments between PAWs and NCPPs persists for the Fe(001) surface. Figure \ref{fig:magmom} gives the magnetic moments as function of layer for a Fe(001) slab. The magnetic moment of a surface atom is $\sim 3$ $\mu_B$ and the difference between the VASP and the SIESTA results is about 3\%. It is well known that $d$-band narrowing for surface atoms enhances the exchange splitting, resulting in a larger magnetic moment for the surface atoms as compared to bulk.\cite{Izquierdo00}

\begin{figure}[t]
\includegraphics[width=8.5cm]{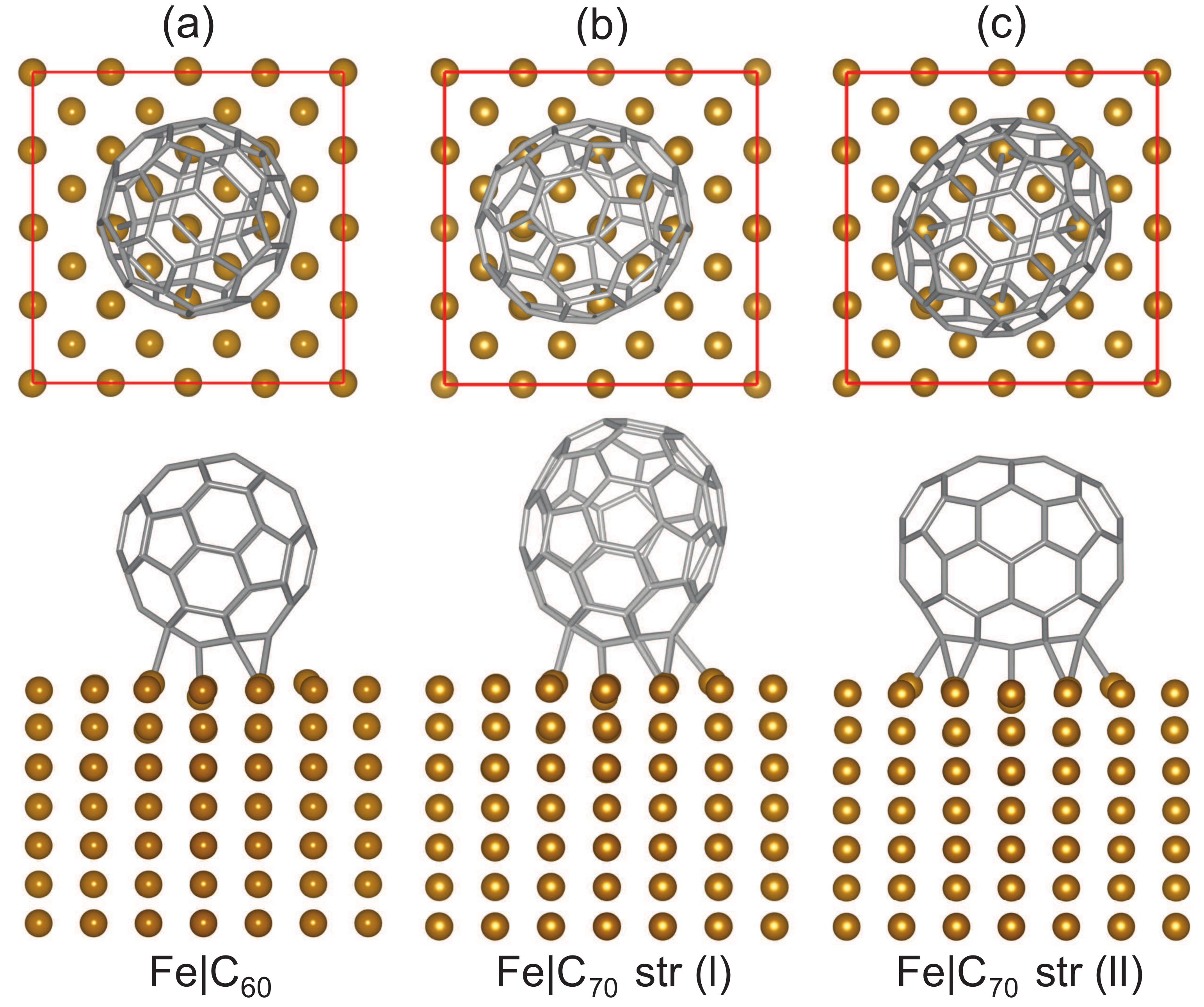}
\caption{(color online) (a) top and side views of the most stable adsorption geometry of C$_{60}$ on Fe(001); Fe--C distances below 2.5 \AA\ are indicated specifically. C$_{70}$ on Fe(001) in structure (I) (b) and structure (II) (c). }
\label{fig:structures}
\end{figure}

\begin{table}[t]
\caption{Binding energies $E_b$ of C$_{60}$ and C$_{70}$ molecules on Fe(001) of the structures shown in Fig.~\ref{fig:structures} (total energies of unperturbed Fe(001) and isolated fullerene minus total energy of fullerene adsorbed on Fe(001)); work function $W$ of Fe(001) covered by a monolayer of fullerenes; magnetic moment $\mu$ induced on the fullerene molecules; spin polarization SP of the density of states at the Fermi level, projected on the fullerene molecules.
}
\begin{ruledtabular}
\begin{tabular}{lcccc}
 structure & $E_b$ (eV) & $W$ (eV)$^{\rm a}$ & $\mu$ ($\mu_B$) & SP (\%) \\
 \hline
C$_{60}$     &   2.94     & 4.81     &  0.22      & 0     \\
C$_{70}$(I)     &   2.79     & 4.67     &  0.22      & 0     \\
C$_{70}$(II)     &   2.99     & 4.79     &  0.22      & 40     \\
\end{tabular}
\end{ruledtabular} 
\newline
$\rm ^a$ Calculated work function of clean Fe(001) is 3.87 eV.
\label{tab:struc}
\end{table}

To model the adsorption of C$_{60}$ and C$_{70}$ molecules we use a slab of seven atomic layers for the Fe(001) substrate with a layer of molecules absorbed on one side of the slab, and 15 \AA\ of vacuum thickness. The molecules and the uppermost three Fe atomic layers are  allowed to relax. A dipole correction is applied to prevent spurious interactions between repeated images of the slab.\cite{Neugebauer92}

From a number of possible adsorption geometries, we have identified the structure of adsorbed C$_{60}$ molecules as most stable that is shown in Fig.~\ref{fig:structures}(a). The edge shared by two hexagons (a 6:6 bond) is on top of a surface Fe atom, and the C$_{60}$ molecule is tilted such that one of the hexagons is more parallel to the surface. There are several short Fe--C bonds in the range 2.0-2.5 \AA, which is an indication for chemisorption, as is confirmed by the binding energy, see Table \ref{tab:struc}. The C--C bond lengths within these two hexagons are between 1.46 and 1.52 \AA, i.e., somewhat larger than the 6:6 and 5:6 bond lengths of 1.40 and 1.46 \AA\ in an unperturbed C$_{60}$ molecule. Judging from the changes in bond lengths, the interaction with the Fe(001) surface seems to break the conjugation in these two hexagons somewhat. The C--C bonds in the other hexagons and pentagons are hardly perturbed by the adsorption. Upon adsorption of a monolayer of C$_{60}$ molecules, the work function of Fe(001) increases by 0.94 eV. The increase indicates that the C$_{60}$ molecule acts as an electron acceptor which is consistent with the high electron affinity of ~4.5 eV of this molecule. The interaction with the ferromagnetic Fe surface induces a small magnetic moment of 0.22 $\mu_B$ on the C$_{60}$ molecule.\cite{Tran13} 

One can form a bonding geometry of the C$_{70}$ molecule to the Fe(001) surface that is very similar to that of C$_{60}$, see Fig. \ref{fig:structures}(b). This structure (I) has the edge shared by two hexagons on top of a surface Fe atom, and the C$_{70}$ molecule is tilted such that one of the hexagons is more parallel to the surface.  C$_{70}$ in this structure has similar properties as C$_{60}$, see Table \ref{tab:struc}, but it is not the lowest energy structure. We find that in the most stable adsorption geometry, structure (II), the long axis of the C$_{70}$ molecule is parallel to the surface, see Fig. \ref{fig:structures}(c). Like in structure (I), in structure (II) the edge shared by two hexagons is on top of a surface Fe atom, but unlike structure (I) the two hexagons in C$_{70}$ have a symmetric tilt with respect to the surface. Again, there are several short Fe--C bonds in the range 2.0-2.3 \AA. The C--C distances in the two hexagons involved in the adsorption are in the range 1.45-1.50 \AA, again somewhat larger than the 1.39-1.47 and 1.44-1.45 \AA\ of the 6:6 and 5:6 bonds, respectively, of the isolated C$_{70}$ molecule, whereas the bond lengths in the rest of the molecule are hardly changed. As structure (II) is 0.2 eV/C$_{70}$ molecule more stable than structure (I), we have used structure (II) in all our transport calculations.     

\begin{figure}[t]
\includegraphics[width=7.5cm]{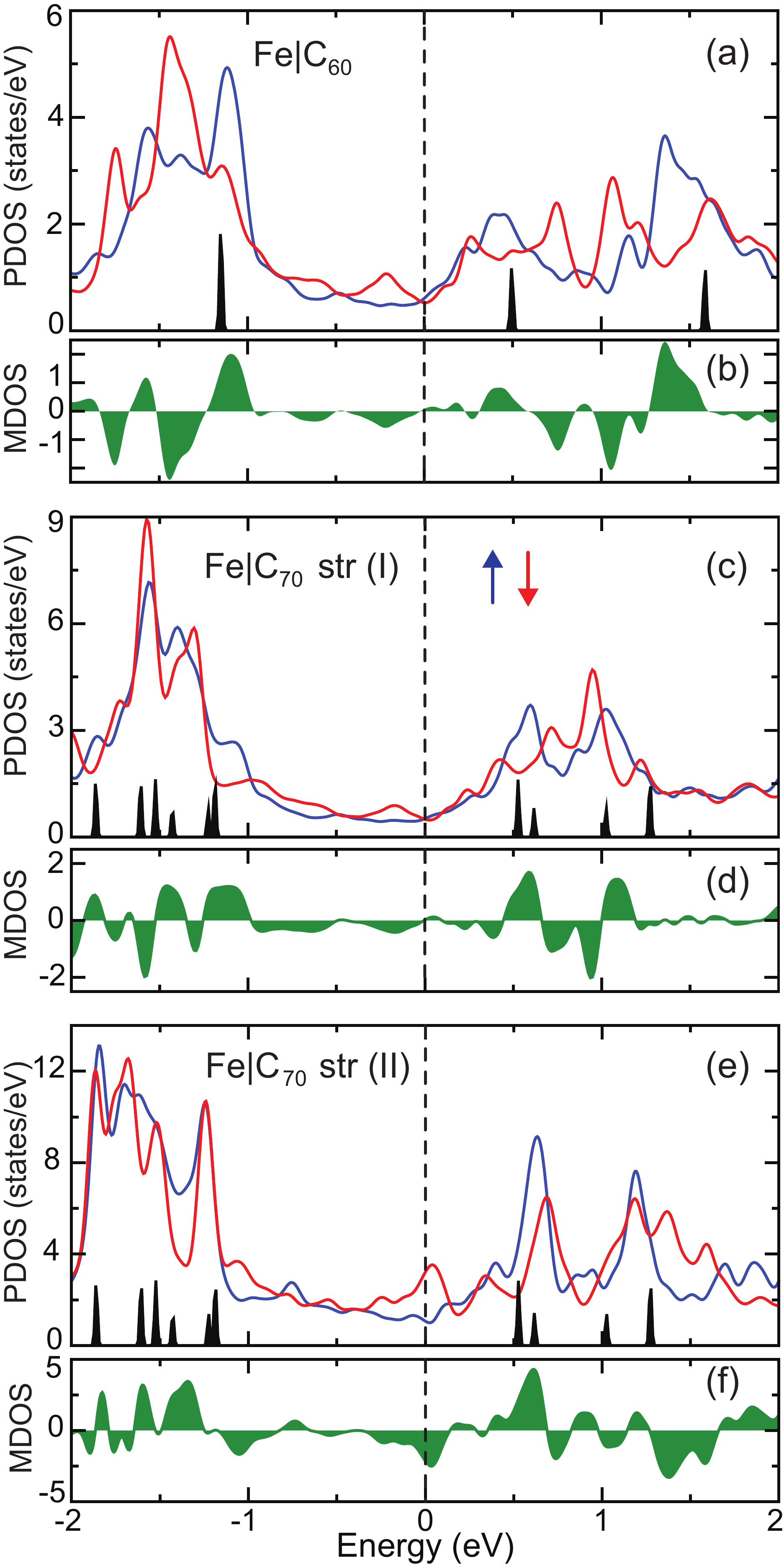}
\caption{(color online) (a) Projected density of states PDOS $n^\uparrow$ of majority (blue) and $n^\downarrow$ of minority (red) spin states of the Fe(001)$|$C$_{60}$ interface, summed over all carbon atoms. Gaussian smearing with a smearing parameter of 0.05 eV is applied. The black lines give the energy levels of the isolated C$_{60}$ molecule. (b) Magnetization density of states  MDOS $\Delta n = n^\uparrow - n^\downarrow$; (c,d) PDOS and MDOS of Fe(001)$|$C$_{70}$ in structure (I), and (e,f) in structure (II).}
\label{fig:DOS}
\end{figure}

The magnetic moments induced on the C$_{60}$ and C$_{70}$ molecules are similar, and do not depend very much on the details of the structure, see Table \ref{tab:struc}. Of course only the spin polarization of the states around the Fermi level is important when studying electron transport, and not the overall polarization or magnetic moment. Figure \ref{fig:DOS}(a) gives the projected density of states (PDOS) of the Fe(001)$|$C$_{60}$ interface, summed over all the carbon atoms. For comparison the Kohn-Sham levels of the isolated C$_{60}$ molecule are also given, which can be aligned with the interface DOS using the lowest $\sigma$-states of the C$_{60}$ molecule. The latter do not participate in the bonding to the surface, and are therefore not perturbed. 

The $\pi$-states of the molecule however hybridize with states from the substrate. These molecular states can still be identified from the peaks in the PDOS, but the peaks are significantly broadened and shifted, compared to the isolated molecule. The isocahedral symmetry $I_h$ of the C$_{60}$ molecule is broken by adsorption on the Fe surface, which lifts degeneracies and splits up the peaks of the adsorbed C$_{60}$ molecule. In addition, the Fe(001) substrate interacts differently with the molecule for different spin states. The Fe(001) surface has prominent surface resonances in the minority spin channel for energies close to the Fermi level.\cite{Stroscio95} The corresponding wave functions have a relatively long decay length, and one can expect these states to interact strongly with adsorbants. Indeed the minority spin states in the PDOS show a stronger perturbation with respect to the molecular $\pi$-states than the majority spin states, in particular for energies around the Fermi level.

\begin{figure}[tb!]
\includegraphics[width=8.0cm]{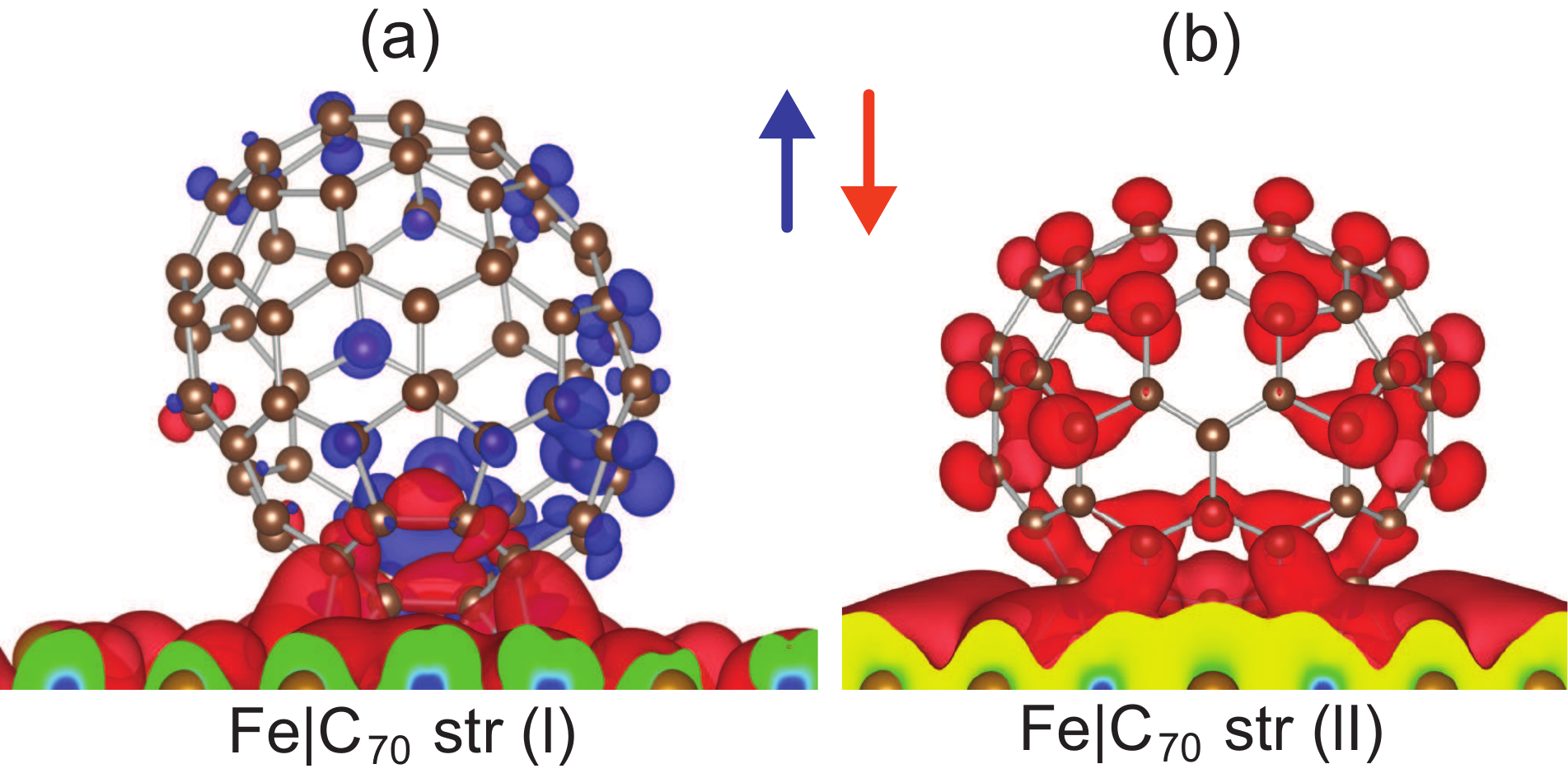}
\caption{(color online) (a,b) Spin polarization of the LDOS at the Fe(001)$|$C$_{70}$ interface in structure I respectively structure II, integrated over an energy interval $[E_F-0.01,E_F+0.01]$ eV.}
\label{fig:LDOS}
\end{figure}

Comparison to the states of the isolated C$_{60}$ molecule allows one to label the corresponding peaks in PDOS of the adsorbed molecule. Of course adsorption broadens the peaks, and sometimes splits them. For instance, the fivefold degeneracy of the molecular HOMO is clearly lifted. In the minority spin states the LUMO, as well as the LUMO+1, which are both threefold degenerate in the isolated molecule, are split up. One of the states derived from the LUMO results in a peak in the minority spin DOS at $E_F-0.2$ eV, whereas other LUMO derived peaks appear above $E_F+0.2$ eV. It indicates that adsorption results in a net transfer of electrons to the C$_{60}$ molecule, which is consistent with an increase of the work function, see Table~\ref{tab:struc}. At the Fermi level, $E = E_F$, the PDOSs of majority and minority spins are nearly equal, which implies that the spin polarization $\Delta n = n^\uparrow - n^\downarrow \approx 0$, see Fig. \ref{fig:DOS}(b). 

Figure \ref{fig:DOS}(c) gives the PDOS of the Fe(001)$|$C$_{70}$ interface, with the C$_{70}$ in structure (I). Again for comparison the Kohn-Sham levels of the isolated C$_{70}$ molecule are also given. The level spectrum of C$_{70}$ is somewhat denser than that of C$_{60}$, as the molecule is slightly larger and less symmetric. Nevertheless, qualitatively the PDOS is remarkably similar to that of C$_{60}$. Specifically, also for C$_{70}$ in structure (I) one of the LUMO-derived states gives a peak in the minority spin at $E_F-0.2$ eV, and other LUMO-derived peaks appear above $E_F+0.2$ eV.  

Figure \ref{fig:DOS}(e) gives the PDOS of the Fe(001)$|$C$_{70}$ interface, with the C$_{70}$ in structure (II). Although qualitatively this PDOS is similar to that of  C$_{70}$ in structure (I), there are nevertheless important differences, specifically for energies around the Fermi level. For C$_{70}$ in structure (II) a hybrid state with C$_{70}$ LUMO character gives a promiment peak in the minority spin channel that is at the Fermi level, instead of $0.2$ eV below $E_F$, as is the case for C$_{70}$ in structure (I) and for C$_{60}$. This means that the spin polarization at $E=E_F$, $\mathrm{SP} = (n^\uparrow - n^\downarrow)/ (n^\uparrow + n^\downarrow) \approx 40$ \% for C$_{70}$ in structure (II), which also implies that the MR in this structure is markedly different, as discussed in Sec.~\ref{sec:results}.

The difference between C$_{60}$ and C$_{70}$ in structure (I) on the one hand, and C$_{70}$ in structure (II) on the other, is also reflected in the wave function at the the Fermi level. Figure \ref{fig:LDOS} shows the spin polarization in the local density of states (LDOS), integrated over an energy interval of $\pm 0.01$~eV around the Fermi level. The LDOS of C$_{70}$ in structure (II) clearly shows shows a hybrid state with clear contributions both from the C$_{70}$ molecule and the Fe(001) substrate, which is delocalized over the whole molecule, and has a clear minority spin character. In contrast, the LDOS of C$_{70}$ in structure (I) shows a hybrid state that covers only part of the molecule, and has a mixed majority/minority spin character. 

                                                                                                                                 
\bibliography{myrefs}

\begin{thebibliography}{71}
\bibitem [{foo()}]{footnoteAP}%
  \BibitemOpen
  \href@noop {} {}\bibinfo {note} {The transmissions of majority and minority
  spins are the same in the AP configuration at zero bias,
  $T_{AP}^\uparrow(E,0)=T_{AP}^\downarrow(E,0)$.}\BibitemShut {Stop}%
\end{thebibliography}

\end{document}